\input amstex

\magnification=1200
\loadbold
\loadeufm
\loadmsbm
\vsize=18.0true cm
\hsize=13.5true cm

\TagsOnRight

\def\sh{\operatorname{sh}}
\def\Ad{\operatorname{Ad}}
\def\ad{\operatorname{ad}}

\def\inv{\operatorname{inv}}

\def\frf#1{\vcenter{\hrule\hbox{\vrule\kern1pt
\vbox{\kern1pt\hbox{$\displaystyle#1$}%
\kern1pt}\kern1pt\vrule}\hrule}}

\def\notni{\hbox{{\kern0pt\raise.7pt\hbox{${\scriptstyle +}$}}
{\kern-9.5pt\raise0pt\hbox{$\supset$}}}}

\def\notnis{\hbox{{\kern0pt\raise.7pt\hbox{${\scriptstyle \times}$}}
{\kern-9.5pt\raise0pt\hbox{$\supset$}}}}

\def\boo{\hbox{{\kern0pt\raise-.2pt\hbox{${}'$}}
{\kern-5.5pt\raise0pt\hbox{$\Bbb I$}}}}

\def\R{\Bbb{R}}
\def\X{\frak X}

\def\notn{\hbox{{\kern0pt\raise.7pt\hbox{${\scriptstyle \times}$}}
{\kern-9.5pt\raise0pt\hbox{$\supset$}}}}

\centerline{\bf INDUCED  REPRESENTATIONS OF CAYLEY--KLEIN}

\medskip
\centerline{\bf  ORTHOGONAL GROUPS}

\bigskip
\centerline{\bf N.A.~Gromov}

\medskip
\centerline{Mathematical Department of Komi Research Centre}

\centerline{of Russian Academy of Sciences,}

\centerline{Kommunisticheskaja str. 24, Syktyvkar}

\centerline{Komi Republic, Russia}

\centerline{E-mail: gromov\@omkomi.intec.ru}

\bigskip
\centerline{\bf S.S.~Moskaliuk}

\medskip
\centerline{Bogolyubov Institute for Theoretical Physics}

\centerline{of National Academy of Sciences of Ukraine}

\centerline{Metrolohichna str. 14-b,}

\centerline{252143, Kyiv-143, Ukraine}

\centerline{E-mail: mss\@gluk.apc.org}

\vskip2cm
 \centerline{\bf Abstract}

\medskip
 Using method of inducing, irreducible unitary representation of
Cay\-ley--Klein orthogonal groups were constructed. There was proved
 that Kirilov's method of orbits is relevant for study of the behavior
 of irreducible  representations under transitions between Cayley--Klein
 groups.

\vfill\eject

{\bf 1. Parameterization and invariant measure}
\medskip

Special orthogonal group $SO(n+1;\boldkey j)$ has been defined in
[1]. Expansion in algebra $so\,(n+1;\boldkey j)$ as a vector
space into direct sum $$
so\,(n+1;\boldkey j)=N_0(\X_{01},\X_{02},\dots,\X_{0n})\oplus
so\,(n;\boldkey j'),
\tag1
$$
where $\boldkey j'=(j_2,j_3,\dots,j_n)$; $\X_{km}$ are matrix
generators $(\X_{\mu\nu})_{\nu\mu}=1$,
$(\X_{\mu\nu})_{\mu\nu}= -\prod\limits_{m=\mu+1}^\nu
 j_m^2$, is invariant in respect to adjoint representation of
subalgebra $so\,(n;\boldkey j')=\{\X_{km},\ k<m,\ k,m=1,2,\dots,n\}$
(special subalgebra in terminology of [2]), because
$[N_0,so\,(n;\boldkey j')]\subset N_0$. Applying expansion (1) to
subalgebra $so\,(n;\boldkey j')$ and setting this process forth, we
obtain complete expansion of algebra $so\,(n+1;\boldkey j)$:  $$
\gather so\,(n+1,\boldkey
j)=N_0\oplus(N_1\oplus(N_2\oplus\dots\oplus(N_{n-2}\oplus\\ \oplus
so\,(2;j_n)))\dots),\tag2 \endgather
$$ where $N_k=N_k(\X_{k,k+1},\dots,\X_{kn})$.

General element $X\,(\boldkey Q_k)\in N_k$ is as follows
$$
X(\boldkey Q_k)=\sum_{s=k+1}^{n}Q_{ks}\X_{ks}
\tag3
$$
where $Q_{ks}\in \Bbb R$; $\boldkey Q_k=(Q_{k,k+1},\dots,Q_{kn})$,
$k=0,1,\dots,n-2$, $\boldkey Q_{n-1}=\mathbreak=Q_{n-1,n}$;
$X\,(\boldkey Q_{n-1})\in so\,(2;j_n)$.

Exponential mapping brings $X\,(\boldkey Q_k)$ into the element of
group \linebreak $SO(n+1;\boldkey j)$ which can be written as
follows:  $$ \align
s(\boldkey Q_k)=e^{X(Q_k)}=&\left(\matrix I_k&0\\
\vspace{10pt} 0&\cos j_{k+1}Q_k\\
\vspace{10pt} 0&\dfrac{Q_{ks}}{Q_k}\dfrac{1}
{j_{k+1}}\sin j_{k+1}Q_k\endmatrix\right.\\ &\left.\matrix
0\\ \vspace{5pt}
-\biggl(\prod\limits_{m=k+2}^{r}j^2_m\biggr)\dfrac{Q_{kr}}
{Q_k}j_{k+1}\sin j_{k+1}Q_k\\
\vspace{5pt}
\delta_{sr}-\biggl(\prod\limits_{m=k+2}^{r}j^2_m\biggr)\dfrac{Q_{ks}Q_{kr}}
{Q_k^2}(1-\cos j_{k+1}Q_k)\endmatrix\right),\tag4
\endalign
$$
where
$r,s=k+1,k+2,\dots,n$;
$Q_k=\biggl(Q^2_{k,k+1}+\sum\limits_{r=k+2}^{n}Q^2_{kr}
\prod\limits_{m=k+2}^{r}j^2_m\biggr)^{1/2}$, $k=0,1,\dots,n-1$, and,
in addition, $Q_{n-1}=Q_{n-1,n}$.

To expansion (2) of algebra $so\,(n+1;\boldkey j)$ there corresponds
to special expansion of the following group transformations $q\in
SO(n+1;\boldkey j)$:
$$
q\equiv q(\boldkey Q_0,\boldkey Q_1,\dots,\boldkey
Q_{n-1})=\prod_{k=0}^{n-1}s(\boldkey Q_k).
\tag5
$$

Further we shall use this parameterization of group
$SO\,(n+1;\boldkey j)$. To find geometrical sense of special group
parameters $\boldkey Q_k$, let us consider connected component of unit
sphere $S_n(\boldkey j)$ in space $\Bbb R_{n+1}(\boldkey j)$,
determined by equation
$$
S_n(\boldkey j)=\biggl\{\boldkey x\in \R_{n+1}(\boldkey
j)|x_0^2+\sum_{k=1}^{n}x_k^2\prod_{m=1}^{k}j_m^2=1\biggr\}.
\tag6
$$
Group $SO\,(n+1;\boldkey j)$ acts transitively on $S_n(\boldkey j)$,
which enables to realize connected component of a sphere as
factor-space $S_n(\boldkey j)=SO\,(n+\mathbreak+1;\boldkey
j)/SO\,(n;\boldkey j')=\{s\,(\boldkey Q_0)\}$, the latter takes place
in parameterization (5). Subgroup $SO\,(n;\boldkey j')$ is
stationary subgroup for point $\boldkey
F_0=(1,0,...,0)$ on sphere. Acting on $\boldkey F_0$ by
transformation $s\,(\boldkey Q_0)$, we get point $\boldkey
y^0=\mathbreak=s\,(\boldkey Q_0)\boldkey F_0$ on sphere with
coordinates $x_0=\cos j_1Q_0$,
$x_r=\dfrac{Q_{0k}}{Q_0}\dfrac{1}{j_1}\times\mathbreak\times\sin
j_1\,Q_0$, $r=1,2,\dots,n$, i.e. $pr_0\boldkey y^0=\boldkey
Q_0\dfrac{\sin j_1Q_0}{j_1Q_0}$, where $pr_0$ is projection on
subspace $\Bbb R_n(\boldkey j')$ along axis $x_0$. We find from here
that $s\,(\boldkey Q_0)$ is a rotation in the plane $\{x_0\boldkey
Q_0\}$, $\boldkey Q_0\in\Bbb R_n(\boldkey j')$ by angle $Q_0$. When
$\boldkey y^0$ runs through all points on sphere, vector $pr_0\boldkey
y^0$ fills all projection of sphere on subspace $\Bbb R_n(\boldkey
j')$. For this reason the range of definition $D_n(\boldkey j)$ of
parameter $\boldkey Q_0$ up to a factor is projection of sphere on
$\Bbb R_n(\boldkey j')$:  $$ D_n(\boldkey j)=\left\{\matrix
\format\l&\quad\l\\ B_\pi^n(\boldkey j')=\{\boldkey Q_0\in
\R_n(\boldkey j')\,|\,Q_0^2\leq \pi^2\},&j_1=1,\\ \vspace{3pt}
\R_n(\boldkey j'),&j_1=\iota_1,\\ \vspace{3pt} B_{i\pi}^n(\boldkey
j')=\{\boldkey Q_0\in\R_n(\boldkey
j')\,|\,Q_0^2\geq-\pi^2\},&j_1=i,\endmatrix\right.  \tag7 $$ where
$B_\pi^n(\boldkey j')$ is a solid sphere of real radius $\pi$ in $\Bbb
R_n(\boldkey j')$; $B_{i\pi}^n(\boldkey j')$ is a solid sphere of
imaginary radius $i\pi$ in $\Bbb R_n(\boldkey j')$.

It is necessary to keep in mind that for some values of parameters
$\boldkey j'$ the sphere of imaginary radius coincides with the whole
$\Bbb R_n(\boldkey j')$, for example, for $j_1=i$,
$j_2=j_3=\dots=j_n=1$, $Q_0^2=\sum_{r=1}^{n}Q_{0r}^2\geq-\pi^2$ for
any $\boldkey Q_0\in\Bbb R_n(\boldkey j')$.

Repeating the same considerations for sphere $S_{n-k}(\boldkey
j^{(k)})$, $k=1,2,\mathbreak\dots,n-2$, $\boldkey
j^{(k)}=(j_{k+1},j_{k+2},\dots,j_n)$, we find that $s\,(\boldkey Q_k)$
is a rotation in the plane $\{x_k,\boldkey Q_k\}$, $\boldkey
Q_k\in\Bbb R_{n-k}(\boldkey j^{(k+1)})$ by angle $Q_k$, where
$\boldkey Q_k<D_{n-k}(\boldkey j^{(k)})$ and $$
\gathered D_{n-k}(\boldkey j^{(k)})=\\
=\left\{\matrix \format\l&\quad\l\\ B_\pi^{n-k}(\boldkey
j^{(k+1)})=\{\boldkey Q_k\in \R_{n-k}(\boldkey
j^{(k+1)})\,|\,Q_k^2\leq \pi^2\},&j_{k+1}=1,\\ \vspace{3pt}
\R_{n-k}(\boldkey j^{(k+1)}),&j_{k+1}=\iota_{k+1},\\ \vspace{3pt}
B_{i\pi}^{n-k}(\boldkey j^{(k+1)})=\{\boldkey Q_k\in\R_{n-k}(\boldkey
j^{(k+1)})\,|\,Q_k^2\geq-\pi^2\},&j_{k+1}=i.\endmatrix\right.
\endgathered
\tag8
$$
For $k=n-1 $ transformation $s\,(\boldkey Q_{n-1})=s\,(Q_{n-1,n})$
is a rotation in the plane $\{x_{n-1},x_n\}$ by angle $Q_{n-1,n}$,
where $Q_{n-1,n}\in D_1(j_n)$, where $$
D_1(j_n)=\left\{\matrix    \format\l&\quad\l\\
[0,2\pi),&j_n=1,\\\R,&j_n=\iota_n,i.\endmatrix\right.
\tag9
$$
Thus, under expression of elements of
group $SO\,(n+1;\boldkey j)$ into product (5) special group
parameters $\boldkey Q_k$ belong to domains $D_{n-k}(\boldkey
j^{(k)})$, described by (7)--(9).

Let us find expression for invariant measure on group
$SO\,(n+1;\boldkey j)$ in parameterization (5). Previously let us
establish what is invariant measure on spheres.  Invariant measure on
sphere $S_n(\boldkey j)$ in Cartesian coordinates is known:
$dF_n(\boldkey x,\boldkey j)=d^nx/x_0$, where
$x_0=\biggl(1-\sum\limits_{k=1}^{n}x_k^2\prod\limits_{m=1}^{k}
j_m^2\biggr)^{1/2}$; $d^nx=dx_1,\dots,dx_n$.
The relation between Cartesian coordinates and parameters $\boldkey
Q_0$ can be found from equation $\boldkey x=s\,(\boldkey
Q_0)\,\boldkey F_0$ and is as follows: $x_0=\cos j_1Q_0$,
$x_r=\dfrac{Q_{0r}}{Q_0}\dfrac{1}{j_1}\sin Q_0j_1$, $r=1,2,\dots,n$.
Jacobian of the transformation is as follows:
$\left|\dfrac{\partial\,(\boldkey x)}{\partial\,(\boldkey
Q_0)}\right|=\left(\dfrac{\sin j_1Q_0}{j_1Q_0}\right)^{n-1}|\cos
j_1Q_0|$, and invariant measure on sphere $S_n(\boldkey j)$ under
parameterization $\boldkey Q_0$ can be written as
$$
dF_n(\boldkey Q_0,j)=\biggl({\sin j_1Q_0\over
j_1Q_0}\biggr)^{n-1}d^nQ_0,
\tag10
$$
where $d^nQ_0=dQ_{01}\dots dQ_{0n}$.

Similarly we find measure on sphere $S_{n-k}(\boldkey j^{(k)})$:
$$
dF_{n-k}(\boldkey Q_k,\boldkey j^{(k)})=\biggl({\sin
j_{k+1}Q_k\over j_{k+1}Q_k}\biggr)^{n-k-1}d^{n-k}Q_k,
\tag11
$$
where $d^{n-k}Q_k=dQ_{k,k+1}\dots dQ_{kn}$.

For some values of parameters $\boldkey j$ the squared quantity
$Q_k^2$ can be positive, negative or zero. Let $D^+=\{\boldkey
Q_k\in D\,|\,Q_k^2>0\}$, $D^-=\{\boldkey Q_k\in\mathbreak\in
D\,|\,Q_k^2<0\}$, $D^0=\{\boldkey Q_k\in D\,|\,Q_k^2=0\}$, then
$D_{n-k}(\boldkey j^{(k)})=D_{n-k}^+(\boldkey j^{(k)})\cup\mathbreak
\cup D_{n-k}^-(\boldkey j^{(k)})\cup D_{n-k}^0(\boldkey j^{(k)})$. In
the domain $D_{n-k}^+(\boldkey j^{(k)})$ invariant measure \linebreak
$dF^+_{n-k}(\boldkey Q_k,\boldkey j^{(k)})$ is given by (11). In the
domain $D_{n-k}^-(\boldkey j^{(k)})$ we have $Q_k=i\widetilde{Q}_k$,
$\widetilde{Q}_k\in\Bbb R$, and for measure we get
$$
dF_{n-k}^-(\boldkey Q_k,\boldkey j^{(k)})=\biggl({\sh
j_{k+1}\widetilde{Q}_k\over
j_{k+1}\widetilde{Q}_k}\biggr)^{n-k-1}d^{n-k}Q_k,
\tag12
$$
The set $D^0_{n-k}(\boldkey j^{(k)})$ has
dimension $n-k-1$ and its measure is equal to zero.

Invariant measure on group $SO\,(n+1;\boldkey j)$ in parameterization
(5) can be written as
$$
dq(\boldkey Q_0,\boldkey Q_1,\dots,\boldkey
Q_{n-1})=\prod_{k=0}^{n-1}
\biggl({\sin j_{k+1}Q_k\over
j_{k+1}Q_k}\biggr)^{n-k-1}d^{n-k}Q_k.
\tag13
$$
It can be shown that this measure is
bilaterally invariant.
\bigskip

{\bf 2. Adjoint algebra, adjoint group,
co-adjoint representation}
\medskip

Further we consider groups $SO\,(n+1;\iota_1,\boldkey j')$ isomorphic
to groups of motions of Cayley-Klein  space of  zero curvature.  To
this  aim we  introduce notations: $Q_{0r}\equiv x_r$,
$r=1,2,\dots,n$, $\boldkey Q_0\equiv \boldkey x$, $s\,(\boldkey
Q_0)\equiv t\,(\boldkey x)=\pmatrix 1&0\\\boldkey x &\boo_n\endpmatrix$,
$\boldkey x\in\Bbb R_n(\boldkey j')$. A set of  transformations
$\{t\,(\boldkey x)\}=N\,(\boldkey x)$ makes Abelian group, and group
$SO\,(n+1;\iota_1,\boldkey j')$ is semidirect product
$$
SO(n+1;\,\iota_1,\boldkey j')=N(\boldkey x)\notnis SO(n;\boldkey
j').
\tag14
$$
The expansion
(5) becomes $g\,(\boldkey x,\boldkey Q_1,\dots,\boldkey
Q_{n-1})=t\,(\boldkey x)\prod\limits_{k=1}^{n-1}s\,(\boldkey Q_k)$;
invariant measure on     the     group     comes     out     of
(13)     for   $j_1=\iota_1$:
$$dg(\boldkey x,\boldkey Q_1,\dots,\boldkey
Q_{n-1})=d^nx\prod_{k=1}^{n-1}
\biggl({\sin j_{k+1}Q_k\over
j_{k+1}Q_k}\biggr)^{n-k-1}d^{n-k}Q_k.
\tag15
$$

Adjoint algebra $\ad L$ of Lie algebra $L$ is defined as $\ad
X\,(Y)=[X,Y]$, $X,Y\in L$, or in matrix form $(\ad
X)_{km}=\sum\limits_{\lambda}c_{\lambda m}^k y_\lambda$, where
$X=\sum\limits_{\lambda}y_\lambda X_\lambda\in\mathbreak\in L$,
$c_{\lambda m}^{k}$ are structure constants of algebra $L$ in a
certain basis. If $Y'=\mathbreak=\ad X\,(Y)$,
$Y=\sum\limits_{\lambda}y_\lambda X_\lambda$,
$Y'=\sum\limits_{\lambda}y'_\lambda X_\lambda$, then relation
$y'_\lambda=\mathbreak=\sum\limits_{m}(\ad X)_{\lambda
m}y_m$ gives in matrix form the action of adjoint algebra on elements
of $L$.

Let us find matrices of adjoint algebra $\ad
(so\,(n+1;\iota_1,\boldkey j'))$ in basis $\X_{km}$. To this aim we
draw up generators $\X_{km}$, $k<m$, in order of increasing of number
$\lambda=m+k\,(n-1)-k\,(k-1)/2$, $\lambda=1,2,\dots,n\,(n+1)/2$ for
$k<m$, $k,m=0,1,\dots,n$. From commutation relations
$$
[X_{\mu_1\nu_1}, X_{\mu_2\nu_2}]=
\cases
\biggl(\prod\limits_{m=\mu_1+1}^{\nu_1} j_m^2\biggr)
X_{\nu_1\nu_2},\quad \mu_1=\mu_2,\quad \nu_1<\nu_2,\\
\biggl(\prod\limits_{m=\mu_2+1}^{\nu_2} j_m^2\biggr)
X_{\mu_1\mu_2},\quad \mu_1<\mu_2,\quad \nu_1=\nu_2,\\
-X_{\mu_1\nu_2},\quad \mu_1<\mu_2=\nu_1<\nu_2.
\endcases
$$

It can
be established that for $k\geq1$ matrices of adjoint algebra of
elements $X\,(\boldkey Q_k)$, given by (3), has block diagonal
structure $\ad X\,(\boldkey Q_k)=\mathbreak=\pmatrix
A_n&0\\0&A_{n(n-1)/2}\endpmatrix$. Let us consider only diagonal block
$A_n$, i.e. the part acting on commutative subalgebra
$N_0(\X_{01},\dots,\X_{0n})$, keeping for this block the notation
$\ad X\,(Q_k)$. We have $\ad X\,(\boldkey x)=0$, $\ad X\,(\boldkey
Q_k)=\mathbreak=\widetilde{X}\,((\boldkey Q_k),\boldkey j^{(k)})$,
$k=1,2,\dots,n-1$, where matrix $\widetilde{X}\,((\boldkey
Q_k),\boldkey j^{(k)})$, comes out of matrix $X\,(\boldkey Q_k)$ by
deleting zero first row and zero first column.

Adjoint group $\Ad (g_0)$ of Lie group $G=\exp L$, $g_0=\exp
X\,(\boldkey F_0)$, is connected with adjoint algebra $\ad
X\,(\boldkey F_0)$ by relation $\Ad (g_0)=\mathbreak=\exp
(\ad X\,(\boldkey F_0))$, and its action on elements $g=\exp
X\,(\boldkey F)$ of group $G$ is described by formula
$$
\Ad\,(g_0)g\overset\text{def}\to=g_0gg_0^{-1}=\exp
X(\Ad\,(g_0)\boldkey F).\tag16
$$
In the case of group $SO\,(n+1;\iota_1,\boldkey j')$ we consider only
that part of matrix of adjoint group which acts on subgroup
$N\,(\boldkey x)$. It is possible due to block diagonal structure of
matrices of adjoint algebra. Under such condition $\Ad
(t(\boldkey x))\equiv I$, and $\Ad (s(\boldkey
Q_k))=\widetilde{s}\,(\boldkey Q_k)$, $k=1,2,\dots,n-1$, where
matrices $\widetilde{s}\,(\boldkey Q_k)$ come out of matrices
$s\,(\boldkey Q_k)$, which are described by (4), by deleting the
first column and the first row, i.e. substituting $I_k$ for $I_{k-1}$.
To the expansion $g= t\,(\boldkey
x)\,\prod\limits_{k=1}^{n-1}s\,(\boldkey Q_k)$ of element of group
$SO\,(n+1;\iota_1,\boldkey j')$ there corresponds the expansion of
matrix of adjoint group:
$$\Ad\,(g)=\Ad\,\biggl(t(\boldkey x)\prod_{k=1}^{n-1}s(\boldkey
Q_k)\biggr)=\prod_{k=1}^{n-1}\Ad\,(s(\boldkey Q_k))=\prod_{k=1}^{n-1}
\widetilde{s}(\boldkey Q_k).\tag17$$

Unitary irreducible representations (characters) of Abelian group of
translations $N\,(\boldkey x)$ are one-dimensional. Each character can
be written as follows
$$H(t(\boldkey x))=\exp(i\langle\boldkey h,\boldkey
x\rangle)=\exp\biggl(i\sum_{k=1}^{n}h_kx_k\biggr),\tag18$$
where $h_k\in \Bbb R$, and group of characters
$\widehat{N}\,(\boldkey h)$ is isomorphic to $N\,(\boldkey x)$. The
action of subgroup $SO\,(n;\boldkey j')$ on $\widehat{N}\,(\boldkey
h)$ is given by
$$
kH(t(\boldkey x))\overset\text{def}\to=H(k^{-1}t(\boldkey
x)k)=\exp(i,\langle\boldkey h,\Ad\,(k^{-1})\boldkey
x\rangle),
\tag19
$$
where $k\in SO_n(\boldkey j')$.

If adjoint group $\Ad(k)$ acts in $N\,(\boldkey x)$ according to the
rule $\boldkey x'=\mathbreak=\Ad(k)\,\boldkey x$, then co-adjoint
group (co-adjoint representation) $\Ad^*(k)$ acts in space of
characters $\widehat{N}\,(\boldkey h)$ according to the rule $\boldkey
h'=\Ad^*(k)\,\boldkey h$, and this action can be found from relation
$\langle\Ad^*(k)\,\boldkey h,\boldkey
x\rangle=\langle\boldkey h,\Ad(k^{-1})\boldkey x\rangle$. In matrix
realization the last requirement gives $\Ad^*(k)=[\Ad(k^{-1})]^T$.
Because $s^{-1}(\boldkey Q_k)=s\,(-\boldkey Q_k)$, we have
$\Ad^*(s\,(\boldkey Q_k))=[\widetilde{s}\,(-\boldkey Q_k)]^T$, or in
explicit form,
$$
\align
\Ad^*(s(\boldkey Q_k))=&\left(\matrix
I_{k-1}&0\\
\vspace{8pt}
0&\cos j_{k+1}Q_k\\
\vspace{8pt}
0&\biggl(\prod\limits_{m=k+2}^{r}j_m^2\biggr)\dfrac{Q_{kr}}
{Q_k}j_{k+1}
\sin j_{k+1}Q_k\endmatrix\right.\\
&\left.\matrix
0\\
\vspace{5pt}
-\dfrac{Q_{ks}}{Q_k}\dfrac{1}{j_{k+1}}\sin j_{k+1}Q_k\\
\vspace{5pt}
\delta_{rs}-\biggl(\prod\limits_{m=k+2}^{r}j^2_m\biggr)\dfrac{Q_{kr}Q_{ks}}
{Q_{k}^2}(1-\cos
j_{k+1}Q_k)\endmatrix\right).\tag20\endalign
$$
To the expansion (5) of elements of group
$SO\,(n+1;\iota_1,\boldkey j')$ there corresponds expansion of
co-adjoint representation
$$\Ad^*(g)=\Ad^*\biggl(t(x)\prod_{k=1}^{n-1}s(\boldkey
Q_k)\biggr)=\prod_{k=1}^{n-1}\Ad^*(s(\boldkey Q_k)).\tag21$$

\bigskip
{\bf 3. Orbits in space of characters
and their stationary}

 {\bf \ \ \ \ subgroups}
\medskip

A set of all $\boldkey h=\Ad^*\,(k)\,\boldkey h_0$, when $k$ runs over
all transformations of $SO\,(n;\boldkey j')$, is called orbit
$O\,(\boldkey h_0)$ of character $\boldkey h_0$ in respect to subgroup
$SO\,(n;\boldkey j')$. It is known [3] that in the case of semidirect
products space of characters $\widehat{N}\,(\boldkey h)$ is splitted
into disjoint orbits, i.e. surfaces in $\widehat{N}\,(\boldkey h)$
invariant in respect to action of co-adjoint representation of group
$Ad^*\,(SO\,(n;\boldkey j'))$. For group $SO\,(n+1;\iota_1,\boldkey
j')$ the equation for orbits can be obtained, substituting generators
$\X_{0k}$, $k=1,2,\dots,n$, in Casimir operator of the second order
$C_2\,(\iota_1,\boldkey j')$ for Cartesian coordinate functions $h_k$
in space $\widehat{N}\,(\boldkey h)$. Casimir operator $C_2(\boldkey
j)$ of group $SO\,(n+1;\boldkey j)$  is as follows:
$$
\align
C_2(\boldkey
j)=&\prod_{r=1}^{n}\biggl(\prod_{m=r+1}^{n}j_m^2\biggr)\X_{0r}^2+\\
&+\sum_{\alpha_2>\alpha_1=1}^{n}\biggl(\prod_{m=1}^{\alpha_1}j_m^2\biggr)
\biggl(\prod_{l=1+\alpha_2}^{n}j_l^2\biggr)\X^2_{\alpha_1\alpha_2}.
\tag22
\endalign
$$
For $j_1=\iota_1$ the
second sum in (22) vanishes, and Casimir operator of group
$SO\,(n+1;\iota_1,\boldkey j')$ is reduced to the first summand in
(22). Substituting then $\X_{0r}$ for $h_r$, we obtain the equation
of orbits
$$
\sum_{r=1}^{n-1}\biggl(\prod_{m=r+1}^{n}j_m^2\biggr)h^2_r+h_n^2
=\text{inv}.
\tag23
$$

The invariant in the right side of (23) can be positive, negative
or zero.  Positive values $\inv=R^2>0$ can be taken for any values of
parameters $\boldkey j'$, negative values $\inv=-\rho^2<0$ -- for all
values of $\boldkey j'$ except for $\boldkey j'=\boldkey 1$ and $\boldkey
j'_k=(j_2,\dots,j_{n-k-1},\iota_{n-k},1,\dots,1)$, $k=0,1,\dots,n-2$.
Zero values $\inv=0$ are possible for any values of parameters
$\boldkey j'$. However,  for $\boldkey j'=\boldkey 1$ orbit degenerates into
point $\boldkey o=(0,\dots,0)$; for $\boldkey j'_k$,
$k=0,1,\dots,n-2$, orbit is entirely situated in subspace $\Bbb
R_{n-k-1}\subset\Bbb R_n$, and, consequently, its dimension is less
than $n-1$. In all other cases the dimension of orbit is $n-1$.

Substituting in Euclidean space $\Bbb R_n$ Cartesian coordinates $h_r$
for $h_r\times\mathbreak\times\prod\limits_{m=r+1}^{n}j_m$,
$r=1,2,\dots,n-1$, we obtain space $\widetilde{\Bbb R}_n(\boldkey j')$
with scalar pro\-duct $(\boldkey h,\boldkey h)_{\bold
j'}=\sum\limits_{r=1}^{n-1}\biggl(
\prod\limits_{m=r+1}^{n}j_m^2\biggr)h_r^2+h_n^2$. Then space of
characters $\widehat{N}\,(\boldkey h)=\mathbreak=\widetilde{\Bbb
R}_n(\boldkey j')$, and the orbits are spheres in space
$\widetilde{\Bbb R}_n(\boldkey j')$ of real $R>0$, ima\-ginary
$i\rho,\ \rho>0$, and zero radius. Depending on values of parameters
$\boldkey j'$, spheres of nonzero radius are either connected, or
consist of two connected components.  Let us consider cases of  real
and imaginary radius separately.

Let $\inv=R^2$, $R>0$. Generators of rotations in planes $\{h_k,h_r\}$
of space  $\Bbb R_n$ we denote $Y_{kr}$. These generators are compact,
if under transition from $\Bbb R_n$ to $\widetilde{\Bbb
R}_n(\boldkey j')$ they are multiplied by real number, and noncompact,
if they are multiplied by imaginary or dual number. Let us consider
generators $Y_{rn}$, $r=1,2,\dots,n-1$, of rotations in planes
$\{h_r,h_n\}$. Under the above mentioned transition they are
multiplied by $\prod\limits_{m=r+1}^{n}j_m$. If all these generators
are noncompact, then the orbit consists of two connected components,
differing in sign of $h_k$. But if at least one of these generators is
compact, then the orbit is connected. Positive part  of coordinate
axis intersects each connected orbit  at the point
$M^+=(0,\dots,0,h_n=R)$, and when the orbit consists of two components,
the whole axis intersects one component at the point $M^+$, and the
other -- at the point $M^-=(0,\dots,0,h_n=-R)$. Analysis of products
$\prod\limits_{m=r+1}^{n}j_m$ enables to state the following
proposition.

{\bf Proposition 1.} {\it Orbits (23) of positive radius make one
family, which is characterized by points $M^+$ for $\boldkey
j'=(j_2,\dots,j_k,j_{k+1}=i,j_{k+2}=\mathbreak=1,\dots,j_{n-1}=1,j_n=i)$,
$k=1,2,\dots,n-2$. In other cases orbits make two subfamilies, one of
which is characterized by points $M^+$ and the other -- by points
$M^-$.}

Let $\inv=-\rho^2$, $\rho>0$. If $j_{m+1}=i$ and parameters
$j_{m+2}=\dots=\mathbreak=j_n=1$, then points
$P^+_m=(0,\dots,0,h_m=\rho,0,\dots,0)$ are intersection points of axis
$h_m$ with orbits of imaginary radius. Generators $Y_{rm}$,
$r=\mathbreak=1,2,\dots,m-1$, are multiplied by
$\prod\limits_{l=r+1}^{m}j_l$.  Similarly to the previous case, the
following proposition is valid.

{\bf Proposition 2.} {\it Orbits (23) of imaginary radius make one
family, characterized by points $P_m^+$ for $\boldkey
j'=(j_2,\dots,j_{m-1},j_m=1,j_{m+1}=\mathbreak=i,j_{m+2}=1,\dots,j_n=1)$,
$m=2,3,\dots,n-1$ and $\boldkey
j'=(j_2,\dots,j_r,j_{r+1}=\mathbreak=i,j_{r+2}=1,\dots,j_{m-1}=1,j_m=i,j_{m+1}=
i,j_{m+2}=1,\dots,j_n=1)$, $r=1,2,\dots,m-2$.
In other cases orbits make two subfamilies, one of which is
characterized by points $P^+_m$, and the other -- by points
$P^-_m=\mathbreak=(0,\dots,0,h_m=-\rho,0,\dots,0)$, $\rho>0$.}

The equation (23) for $\inv=0$ gives one orbit and not their family.
For $\boldkey
j'=(j_2,\dots,j_{n-k-1},j_{n-k}=\iota_{n-k},j'_{n-k+1},\dots,j'_s,
j_{s+1}=i,j_{s+2}=\mathbreak=1,\dots,j_n=1)$, where
$j'_{n-k+1},\dots,j'_s=1$, $i,s=n-k,n-k+1,\dots,n-1$,
$k=0,1,\dots,n-2$, orbit of zero radius can be characterized by points
$\widetilde{O}=(0,\dots,0,h_s=\pm1,0,\dots,0,h_n=\pm1)$.

{\bf Definition 1.} {\it Group $K_{\bold h_0}=\{s\in SO\,(n;\boldkey
j')\,|\,\Ad^*(s)\,\boldkey h_0=\boldkey h_0\}$ is called stationary
subgroup $K_{\bold h_0}$ of orbit $O\,(\boldkey h_0)$.}

Using (20), (21) we find for the family of orbits of positive
radius that stationary subgroup of points $M^{\pm}$ is subgroup of
group $SO\,(n;\boldkey j')$, consisting of transformations, leaving
invariant axis $x_n$, i.e.
$K_{M^{\pm}}=\mathbreak=SO\,(n-1;j_2,\dots,j_{n-1})$; moreover,
transformations $s\,(n)\in SO\,(n-\mathbreak-1;j_2,\dots,j_{n-1})$ can
be written as follows:  $$ s(n)=\prod_{k=1}^{n-2}s(\boldkey
Q_k(Q_{kn}=0)).  \tag24 $$

For the family of orbits of imaginary radius stationary subgroup of points
$P^{\pm}_m$ is subgroup of group $SO\,(n;\boldkey j')$, leaving
invariant the axis $x_m$, i.e.
$K_{P^{\pm}_m}=SO^m_{n-1}(j_2,\dots,j_n)$, where
transformation $s\,(m)\in SO^m(n-\mathbreak-1;j_2,\dots,j_n)$ is as
follows
$$
s(m)=\prod_{k=1}^{m-1}s(\boldkey
Q_k(Q_{km}=0))\prod_{r=m+1}^{n-1}s(\boldkey Q_r).
\tag25
$$

{\bf 4. Irreducible unitary representations}
\medskip

Group $K_{\bold h_0}$ is stabilizer of character $\boldkey h_0$ in
respect to action of co-adjoint representation of group
$SO\,(n;\boldkey j')$. Let $T_{\bold h_0}$ be irreducible unitary
representation of group $K_{\bold h_0}$ in some Hilbert space $\Cal
H_T$. Then irreducible unitary representation $e^{i\bold h_0}\otimes
T_{\bold h_0}$ of subgroup $N\,(\boldkey x)\notn K_{\bold h_0}$ is
realized in $\Cal H_T$ by relation
$$
e^{i\bold h_0}\oplus T_{h_0}(t(\boldkey x)s(\boldkey
h_0))=e^{i\langle\bold h_0,\bold x\rangle}T_{\bold
h_0}(s(\boldkey h_0)),
\tag26
$$
where $t\,(\boldkey x)\in N\,(\boldkey x)$; $s\,(\boldkey h_0)\in
K_{\bold h_0}$.

If $\inv=R^2$ then $\boldkey h_0=M^{\pm}$, $\langle\boldkey
h_0,\boldkey x\rangle=\pm Rx_n$, and irreducible unitary
representation of subgroup $N\,(\boldkey x)\notn
SO\,(n-1;j_2,\dots,j_{n-1})$ is \linebreak $e^{\pm
iRx_n}T_{M^{\pm}}(s\,(n))$.  If $\inv=-\rho^2$, then
$\boldkey h_0=P^{pm}_m$, $\langle\boldkey h_0,\boldkey
x\rangle=\pm\rho x_m$ and irreducible unitary representation of
subgroup $N\,(\boldkey x)\notn SO^m\,(n-1;\boldkey j')$ is $e^{i\rho
x_m}T_{P^{\pm}_m}(s\,(m))$.

Each irreducible unitary representation of group
$SO\,(n+1;\iota_1,\boldkey j')=\mathbreak=N\,(\boldkey x)\notn
SO\,(n;\boldkey j')$ is induced by irreducible unitary representation
\linebreak
$e^{i\bold h_0}\otimes T_{\bold h_0}$ of its subgroup $N\,(\boldkey
x)\notn K_{\bold h_0}$. Let us denote these representations  of group
$SO\,(n+1;\iota_1,\boldkey j')$ by symbol $\omega_{\bold h_0,T}$.
Operators $\omega_{\bold h_0,T}$ act in Hilbert space of
square-integrable functions on $SO\,(n;\boldkey j')$
$$
\align\Cal H_{\bold h_0,T}=&\{f\,|\,f(k(\boldkey Q))\in
L^2(SO(n;\boldkey j'))\,\&\,T_{\bold h_0}(s(\boldkey
h_0))f(k(Q)s(\boldkey h_0))=\\
=&f(k(\boldkey Q))\forall s(\boldkey h_0)\in K_{\bold h_0},\quad
k(\boldkey Q)\in SO(n;\boldkey j')\}
\tag27\endalign
$$
according to
the rule
$$
\align
\omega_{\bold h_0,T}&(t(\boldkey x)k(\boldkey Q))f(k(\boldkey
Q^0))=\\&e^{i\langle\bold h_0,\Ad(k^{-1}(\bold Q^0))\bold
x\rangle}f(k^{-1}(\boldkey Q)k(\boldkey Q^0)).
\tag28\endalign
$$

Let us evaluate vector $\boldkey x^{(n-1)}=\Ad\,(k^{-1}(\boldkey
Q))\,\boldkey x$. Group element \linebreak $k\,(\boldkey Q)\in
SO\,(n;\boldkey j')$ in parameterization (5) can be written as product
$k\,(\boldkey Q)=\prod\limits_{k=1}^{n-1}s\,(\boldkey Q_k)$. Then
$k^{-1}(\boldkey Q)=\prod\limits_{k=n-1}^{1}s\,(-\boldkey Q_k)$ and,
correspondingly, $\Ad\,(k^{-1}(\boldkey
Q))=\prod\limits_{k=n-1}^{1}\Ad\,(s\,(-\boldkey Q_k))$. Let us
introduce notations:
$$
\boldkey x^{(k)}=\prod_{r=k}^{1}\Ad(s(-\boldkey Q_r))\boldkey
x,
\tag29
$$
$$
X_k=x_k{1\over j_{k+1}}\sin j_{k+1}Q_k+{(Q_k,\boldkey
x)\over\boldkey  Q_k} (1-\cos j_{k+1}Q_k),
\tag30
$$
$$
\gather
A_{pk}=-{Q_{pk}\over Q_p}{1\over j_{k+1}}\sin j_{k+1}Q_k+
{(\boldkey Q_p,\boldkey Q_k)\over Q_pQ_k}(1-\cos j_{k+1}Q_k),\\
p<k,
\tag31\endgather
$$
$$
Y_k=x_k\cos j_{k+1}Q_k+{(\boldkey Q_k,\boldkey x)\over
Q_k}j_{k+1}\sin j_{k+1}Q_k,
\tag32
$$
$$
\gather
B_{pk}=-{Q_{pk}\over Q_p}\cos j_{k+1}Q_k+
{(\boldkey Q_p,\boldkey Q_k)\over Q_pQ_k}j_{k+1}\sin j_{k+1}Q_k,\\
p<k,
\tag33\endgather
$$
$$
\gather
(\boldkey Q_k,\boldkey
x)=\sum_{r=k+1}^{n}\biggl(\prod_{l=k+2}^{r}j_l^2\biggr)Q_{kr}x_r,\\
(\boldkey Q_p,\boldkey Q_k)=\sum_{r=k+1}^{n}
\biggl(\prod_{l=k+2}^{r}j_l^2\biggr)Q_{pr}Q_{kr},
\tag34\endgather
$$
where $\prod\limits_{l=k}^{r}j_l^2\equiv1$ for $r<k$. Rather
cumbersome calculations give:
$$
\gathered
x_r^{(k)}=x_r^{(r)}=Y_r-\sum_{p=1}^{r-1}D_pB_{pr},\quad
r=1,2,\dots,k,\\
x_r^{(k)}=x_r-\sum_{p=1}^{k-1}D_p{Q_{pr}\over Q_p},\quad
r=k+1,k+2,\dots,n,
\endgathered
\tag35
$$
where coefficients $D_p$ satisfy
recurrent relation
$$
D_p=X_p-\sum_{s=1}^{p-1}D_sA_{sp},\quad D_1=X_1.
\tag36
$$
To find $D_p$, we proceed to the set of linear
equations
$$
D_p+\sum_{s=1}^{p-1}D-sA_{sp}=X_p
\tag37
$$
or in matrix form $A\boldkey D=\boldkey X$, where $A$ is triangle
matrix of dimension $p$ with units on main diagonal; nonzero matrix
elements are $(A)_{rs}=A_{sr}$, $s<r$, $s=1,2,\dots,p-1$,
$r=2,3,\dots,p$. It is obvious that $\det A=1$. By Cramer rule, we
find
$$
D_p=\det\left|\matrix
1&0&0&\dots&0&X_1\\
A_{12}&1&0&\dots&0&X_2\\
A_{13}&A_{23}&1&\dots&0&X_3\\
\vdots&\vdots&\vdots&\ddots&\vdots&\vdots\\
A_{1,p-1}&A_{2,p-1}&A_{3,p-1}&\dots&1&X_{p-1}\\
A_{1,p}&A_{2,p}&A_{3,p}&\dots&A_{p-1,p}&X_{p}\endmatrix\right|
\tag38
$$
Thus coefficients $D_p$  are determined up to evaluation of
determinant of $p$-th order. Knowing $D_p$, one can find from (35)
that
$$
x_n^{(n-1)}=(\Ad(k^{-1}(\boldkey Q^0)\boldkey
x)_n=x_n-\sum_{p=1}^{n-1}D_p^0{Q_{pn}^0\over Q_p^0},
\tag39
$$
$$
x_m^{(n-1)}=x_m^{(m)}=Y_m^0-\sum_{p=1}^{m-1}D_p^0B_{pm}^0.
\tag40
$$

Hence, to family of orbits of real radius $R>0$ there corresponds a
series of irreducible unitary representations of group
$SO\,(n+1;\iota_1,\boldkey j')$, which is realized by operators
$$
\gather
\omega^\pm_{R,T_M}(t(\boldkey x)k(\boldkey Q))f(k(\boldkey Q^0))=\\
=\exp\biggl\{\pm iR\biggl(x_n-\sum_{p=1}^{n-1}D^0_p{Q_{pn}^0\over
Q_p^0} \biggr)\biggr\}f(k^{-1}(\boldkey Q)k(\boldkey
Q^0))\tag41
\endgather
$$
in Hilbert space $\Cal H_{R,T_M}$.

Family of orbits of imaginary radius $i\rho$, $\rho>0$, gives a series
of irreducible unitary representations of group
$SO\,(n+1;\iota_1,\boldkey j')$, which operators
$$
\gather
\sigma^\pm_{\rho,T_{P_m}}(t(\boldkey x)k(\boldkey Q))f(k(\boldkey
Q^0))=\\ =\exp\biggl\{\pm
i\rho\biggl(Y_m^0-\sum_{p=1}^{m-1}D^0_pB_{pm}^0
\biggr)\biggr\}f(k^{-1}(\boldkey Q)k(\boldkey
Q^0))\tag42
\endgather
$$
act in Hilbert space $\Cal H_{\rho,T_{P_m}}$. The sign plus and minus
in (41), (42) are chosen in accordance with Propositions 1 and
2.

Let us consider the simplest case $j'_i=i'_i$,
$j_2=\iota_2,\dots,j_n=\iota_n$. Group
$SO\,(n+1;{\boldsymbol\iota})$ is group  $Z\,(n+1)$ of the lower
triangular matrices of $n+1$-th order with units in the main diagonal
and all elements above the main diagonal equal to zero. For $n=2$
group $SO\,(3;\iota_1,\iota_2)$ is Heisenberg group, for which we,
using (30), (31), (38), (41), obtain irreducible representations
$$
\omega^\pm_{R}(t(\boldkey x)s(Q_{12}))f(s(Q^0_{12}))=
e^{\pm iR(x_2-x_1Q_{12}^0)}f(s(Q_{12}^0-Q_{12} )),
\tag43
$$
coinciding with known representation of this group [4]. For $n\geq3$
we derive from (41) irreducible unitary representations of group
$Z\,(n+1)$ [5]. Comparing, it is necessary to keep in mind that in
the handbook by D.P.~Zhelobenko, A.I.~Stern [5] operators of
representation act on functions depending on other variables that
occurring in (27).

Using the method of inducing, we have constructed irreducible unitary
representations of groups $SO\,(n+1;\iota_1,\boldkey j')$. Can all
these representation be obtained from irreducible representations of
group $SO\,(n+1;\iota_1,\boldkey 1)=\mathbreak=N\,(\boldkey x)\notn
SO\,(n;\boldkey 1)$ of notions of $n$-dimensional Euclidean space?
(Representations of Euclidean group are described by (27), (41),
when all parameters take unit values:  $j_k=1$.) One can, first,
derive only representations $\omega$, corresponding to orbits of real
radius. Second, parameter $R>0$, which numbers irreducible
representations of Euclidean group, has to be substituted for $\pm R$
in accordance with Proposition 1, in order to obtain both series of
irreducible representations, when they exist.

Under transition from Euclidean group to $SO\,(n+1;\iota_1,\boldkey
j')$ subgroup $SO\,(n)$ turns into $SO\,(n;\boldkey j')$. This changes
space of functions (27), namely: the range of definition for
functions changes, therefore the requirement of square-integrability
changes the functions, the structural condition, imposed on functions,
is modified as well, because stabilizer of character $SO\,(n-1)$ goes
into $SO\,(n-1;j_2,\dots,j_{n-1})$.

Operators of representation (41) change less radically. Here
elements $g^*$ of Euclidean group are replaced by elements $g\in
SO\,(n+1;\iota_1,\boldkey j')$, and the exponent $R^*\,(\boldkey
x^*,\boldkey Q^*)$ under transition $\psi$:
$SO\,(n+1;\iota_1,\boldkey 1)\to\mathbreak\to
SO\,(n+1;\iota_1,\boldkey j')$, $\psi x^*_1=x_1$, $\psi
x^*_k=x_k\prod\limits_{l=2}^{k}j_l$, $k=2,3,\dots,n$, $\psi
Q^*_{pk}=\mathbreak=Q_{pk}\prod\limits_{l=p+1}^{k}j_l$, $p<k$,
$p,k=1,2,\dots,n$, acquires factor $\prod\limits_{l=2}^{n}j_l$.  For
this reason it must be transformed according to the rule $R\,(\boldkey
x,\boldkey
Q)=\biggl(\prod\limits_{l=2}^{n}j_l^{-1}\biggr)\times\mathbreak\times
R^*(\psi\boldkey x^*,\psi\boldkey Q^*)$. (All quantities, referring
to Euclidean group are marked with asterisk.)

Because under pure contractions orbits of imaginary radius do not
emerge, we, transforming irreducible representations of Euclidean
group, obtain irreducible representations of contracted group.
Moreover, if $j_r=\mathbreak=\iota_r,j_{r+1}=\dots=j_n=1$, and
parameters $j_m=1,\iota_m,i$, $m=2,3,\dots,r-1$, then such groups do
not have orbits of imaginary radius, and all their irreducible
representations can be obtained, as described above, transforming
irreducible representations of Euclidean group.

Only transitions from Euclidean group similar to analytical
continuation and combination of contractions and analytical
continuation, different from mentioned in previous paragraph, give
irreducible representations of contracted groups, corresponding to
orbits of real radius, and do not allow to obtain representations
corresponding to orbits of imaginary radius. Let us remind that we
have not considered representations generated by orbits of zero
radius. Let us notice as well that  Kirilov's method of orbits [3]
proves to be relevant for study of the behavior of irreducible
representations under transitions between groups.

\medskip
The authors are grateful to Yu.A.~Danilov for helpful discussions.

\bigskip
\centerline{\bf References}

\medskip
\item{1.}
Gromov N.A., Moskaliuk S.S. Special orthogonal groups in Cayley--Klein
spaces // Hadronic Journal. -- 1995, Vol.18, Iss.5.  -- 
P.451--483.

\item{2.}
Zaitsev G.A. Algebraic problems of mathematics and
theoretical\linebreak physics. -- Moscow: Nauka, 1974. -- 192 p.

\item{3.} Kirillov A.A. Elements of theory of representations. --
 Moscow:\linebreak  Nauka, 1978. -- 344 p.

\item{4.}
Barut A., Ronczka P. Theory of group representations. Moscow: Mir,
1980. -- Vol.1 -- 456 p.; Vol.2 -- 396 p.

\item{5.}
Zhelobenko D.P., Stern A.I. Representations of Lie groups. --
 Moscow: Nauka, 1983. -- 360 p.

\end